\documentstyle[prl,multicol,aps,epsf]{revtex}
\begin{document}
\title{Wave Function Mismatches and Coulomb Drag}
\author{Fei Zhou}
\address{Physics Department, Princeton University, Princeton, NJ 08544}
\maketitle

\begin{abstract}
In this paper 
I study the topological excitations
in a pairing state in double layer systems
at $\nu=1/2$ in the presence of disorders. 
Due to mismatches between single particle wave functions 
of composite Fermions in different layers, the sensitivity of the
Chern number of the pairing composite Fermion
state,  with respect to changes of impurities,
is infinity.
Consequently, Goldstone mode in this pairing state is strongly
damped at low temperature.  
This leads to a unique temperature dependence of the drag resistance
at low temperature.
\end{abstract}


\begin{multicols}{2}

Coulomb drag in double layer systems with Landau level filling factor $\nu=1/2$
in each layer has attracted much attention for the past few years.
\cite{eisenstein,kim,stern,kamenev,stern2,zhou99,Girvin,bonesteel}.
When the distance between layers $d$ is 
less than the magnetic length $L_H$, 
numerical result of \cite{Girvin} 
shows that the ground state can be approximated
by a $\{111\}$ state proposed in \cite{Halperin};
when $d \gg L_H$, it was argued that 
the interlayer attractive interaction mediated by 
the Chern Simons field leads to an S-wave pairing BCS state
of composite Fermions at half filled Landau levels\cite{bonesteel}.
This is a generalization of the idea of having an incompressible 
quantum Hall liquid with even-denominator filling factor 
discussed some years ago\cite{Halperin,Haldane,Wen,Read}.
Attempts have been made to identify these incompressible
states proposed in\cite{Halperin,Haldane,Wen,Read}
with pairing composite Fermion states.
As pointed out in \cite{Read99}, $\{331\}$ state in\cite{Halperin},
and Pfaffian states in\cite{Wen,Read}
can be expressed in terms of $p$-wave pairing composite Fermion
states with total spin (in the layer index space in the present case)
equal to zero along different directions;
while the hollow-core state in
\cite{Haldane} can be connected with a singlet $d$-wave pairing.

These considerations suggest that 
low energy transport in double layer systems at $\nu=1/2$, 
in general, is determined by the properties of a pairing 
composite Fermion state. 
Especially, in the absence of tunneling between two layers
and at zero temperature,
the Coulomb drag in a pairing state 
is nonzero\cite{zhou99}. In contrast, the drag between 
two Fermi liquids vanishes as the temperature $T$ goes to 
zero. 
The in-plane resistance $R_{11}$ and the
interlayer drag resistance $R_{12}$ in double layers
in a pairing state of composite Fermions can be expressed as\cite{zhou99}

\begin{equation}
R_{11, 12}=
\frac{1}{2\sigma_{quasi}} \pm \frac{1}{2\sigma_{cond}}.
\end{equation}
For an S-wave pairing state,
$\sigma_{quasi}=g \exp (-\Delta/kT)$ is the conductivity
from thermally excited 
quasiparticles; $g=k_F l \times e^2/\hbar$ and $\Delta$ is the 
quasiparticle energy gap of the pairing state, or the mobility gap
when the quasiparticles are localized. 
$l$ is the elastic
mean free path of the composite Fermions in the
presence of disorders, much longer than the Fermi wave length 
$k^{-1}_F$.  ${\sigma}_{cond}$ is from the composite Fermion
condensate, as defined in Eq.12.

Meanwhile,

\begin{equation}
\frac{\kappa^e(q)}{\kappa^{cF}(q=0)}
=(1+\frac{k_F^2N_s}{4q^2N_e})^{-1}.
\end{equation}
Here $\kappa^{e,CF}$ are the irreducible
compressibility of electrons and 
composite Fermions respectively at wave vector $q$;
$N_{s,e}$ are the composite Fermion superfluid density
and electron number density.
Following Eqs.1,2, $R_{11}=R_{12}$ when 
composite Fermions have an off-diagonal long range order
($N_s\neq 0$, $\sigma_{cond}=\infty$) and 
the double layer system becomes incompressible, i.e, $\kappa^e(0)=0$.

In the following I consider the
topological excitations in
a pairing state of composite Fermions in a disordered double
layer system.
Due to mismatches between single particle wave functions for
composite Fermions in different layers,
the pairing state wave function in the presence of different impurity
configurations is topologically different. Furthermore,
topological excitations have a continuous energy spectrum. 
These low lying topological excitations lead to a novel coupled
double layer liquid, characterized by 

\begin{equation}
R_{11}-R_{12}=\frac{1}{\sigma_{cond}}\propto T^2
\end{equation}
at low temperature limit. 
Eq. 3 is valid only in the strongly coupled 
limit when $\sigma_{cond} \gg \sigma_{quasi}$.

To obtain these results, let me first
consider a double layer 2DEG where
the impurity scattering
potential in one layer $V_1({\bf r})$ is different from
the impurity potential in another layer $V_2({\bf r})$.
The correlation of these two sets of impurity potentials is determined by

\begin{equation}
\theta_{12}=1-\frac{<V_1({\bf r})V_2({\bf r})>}{<V^2_1({\bf r})>}.
\end{equation}
Furthmore, I assume short range impurity
potential $s^{-1}<V_1({\bf r})V_1({\bf r'})>
=s^{-1}<V_1({\bf r})V_1({\bf r'})>
=\delta({\bf
r}-{\bf r'})V_{0}^2$, $s$ is the area of the sample.
When $V_1, V_2$ are completely (un)correlated, $\theta_{12}=(1)0$. 

Let me introduce the composite Fermion pairing wave function
in a general form: $\Delta({\bf r}, {\bf R})=<\psi_1({\bf R}+{\bf r}/2)
\psi_2({\bf R}-{\bf r}/2)>\chi(1,2)$.
$\psi_{1,2}$ are the composite Fermion operators in layer 1,2 and
$\chi(1,2)$ is the pair wave function in layer index space. 
It satisfies 
the gap equation generalized to a double layer system
with an interlayer attractive interaction constant $\lambda_{12}$\cite{abr}, 

\begin{eqnarray}
&&\Delta({\bf r}, {\bf R})=\int d{\bf r'}d{\bf r''}d{\bf R'}
\lambda_{12}({\bf r},{\bf r''}) \int d\epsilon [1-2n_F(\epsilon)]
\Delta({\bf r'},{\bf R'})  
\nonumber \\
&&[ G^{1R}_{0\epsilon}({\bf R}+{\bf r''}/2, {\bf R'}+{\bf r'}/2) 
G^{2A}_{-\epsilon}({\bf R}-{\bf r''}/2, {\bf R'}-{\bf r'}/2)
\nonumber \\
&&+G^{2R}_{0\epsilon}({\bf R}+{\bf r''}/2, {\bf R'}+{\bf r'}/2) 
G^{1A}_{-\epsilon}({\bf R}-{\bf r''}/2, {\bf R'}-{\bf r'}/2) ]
\end{eqnarray}
Here  $n_F(\epsilon)$ is the Fermi distribution function;
$G^{1R, 2R}_{0\epsilon}({\bf r}, {\bf r'})$ 
($G^{1A, 2A}_\epsilon({\bf r}, {\bf r'})$)
are the retarded(advanced) exact Green
function in layer 1 and 2 in the {\em absence}({\em presence}) of pairing
potential. 
Depending on the form of $\lambda_{12}$, the order parameter
would have different symmetries. 
In a clean limit, when $\lambda_{12}({\bf r}, {\bf r'})
=\lambda_s\delta({\bf r})\delta({\bf r'})$, $\Delta({\bf r}, {\bf
R})=
\delta({\bf r})\Delta_s({\bf R})\chi^A(1,2)$ corresponding to an S-wave state.
When $\lambda_{12}({\bf r}, {\bf r'})
=\lambda_p k_F^{-2}\nabla_i\delta({\bf r})\nabla_i\delta({\bf r'})$,
$\Delta({\bf r},
{\bf R})=
\nabla_i\delta({\bf r})\Delta_{pi}({\bf R})\chi^S(1,2)$ 
representing a p-wave state, $i=x,y$,
$\chi^{S}(1,2)=\chi^S(2,1)$ and $\chi^A(1,2)=-\chi^A(2,1)$.
Without losing generality,  
I will restrict myself to the S-wave pairing composite Fermion state.
The final conclusion can be extended to a p-wave pairing state.

Consider the exact Green function at $\epsilon=0$
and $\delta {\bf a}=0$ in $2D$.
$G^{1R,2R}_{00}({\bf r}, {\bf r'})=
\cos(k_F|{\bf r}-{\bf r'}|-\pi/2 +\phi_{1,2}({\bf r}))
\times m/\sqrt{k_F|{\bf r}-{\bf r'}|}$, with $\phi_{1,2}({\bf r})$
as the random phase shift at point ${\bf r}$ in the presence of impurity
potential $V_{1,2}({\bf r})$.
By the same token, in a given
sample, $G^{1R}_{00}({\bf r}, {\bf r'})
G^{2A}_{00}({\bf r}, {\bf r'})\propto |{\bf r}-{\bf r'}|^{-1}
\cos(\phi_1({\bf r})-\phi_2({\bf r}))$,
with the fast oscillating phase factor $k_F|{\bf r}-{\bf r'}|$
cancelled out.
When $V_1({\bf r})$ is not completely correlated with
$V_2({\bf r})$, $\phi_1({\bf r})-\phi_2({\bf r})$ is finite.
$G_{00}^{1R}G_{00}^{2A}$ consists a series of nodes at points ${\bf r}$
where $\phi_1({\bf r})-\phi_2({\bf r})=(n+1/2)\pi$,
due to mismatches of two sets
of single particle wave functions at the Fermi energy in a double layer
system.
{\em However, the presence of these nodes
only reveals itself via an exponential decay of Green functions with a given
sign when the ensemble 
averaged is taken\cite{Spivak}}.
For example,
$\int d\epsilon <G^{1R}_{0\epsilon}({\bf r}, {\bf r'})
G^{2A}_{0-\epsilon}({\bf r}, {\bf r'})>=   
{m}(2\pi |{\bf r}-{\bf r'}|^2)^{-1}
\exp(-{\theta_{12} |{\bf r}-{\bf r'}|}/{l})$,
$<>$ represents the impurity average.
Consequently, following Eq.5, at $T=0$ 
the self consistent equation
doesn't have a homogeneous solution with 
$\delta {\bf a}=0$ and $\Delta_s({\bf r})$ a constant when
$\theta_{12}\geq \theta_c=\tau\Delta_0(\ll 1)$;
$\Delta_0$ is the order parameter in the absence of
disorders and $\tau$ is the elastic scattering time.

To account {\em random signs} of Green functions in the presence
of mismatches of wave function, 
one has to take into account higher order contributions
in term of $(k_Fl)^{-1}$; this is beyond 
the conventional approximation used in the theory of disordered
superconductors\cite{abr}.
Performing an expansion of
the self consistent equation around $\theta_c$ 
diagrammatically (See Fig.1), I obtain for $\Delta_s$,

\begin{eqnarray}
&&[\xi^2_0\tilde{\nabla}^2 +
\frac{\delta\theta_{12}}{\theta_c}]\Delta_s({\bf
r})+ \int d{\bf r} K_s({\bf r}, {\bf r'})\Delta_s({\bf r'})=
\frac{|\Delta_s|^2}{\Delta_0^2}\Delta_s({\bf r}),
\nonumber \\
&&<K_s({\bf r}, {\bf r'})>=0,\nonumber \\
&&<K_s({\bf r_1}, {\bf r'_1}) K_s({\bf r_2}, {\bf r_2'})>=
\nonumber \\
&&\frac{1}{16 g^2}[\delta({\bf r_1}-{\bf r_2})\delta({\bf r'_1}-{\bf r'_2})
[\delta({\bf r_1}-{\bf r'_1})
+ \frac{\xi_0^2}{|{\bf r_1}-{\bf r'_1}|^4}].
\end{eqnarray}

Here $\tilde{\nabla}=\nabla-i\delta {\bf a}$,
$\delta \theta_{12}=\theta_{12}-\theta_c$,
$\xi_0=\sqrt{D/\Delta_0}$.  Meanwhile,
$\delta {\bf a}={\bf a} +{\bf A}$,
${\bf a}$ is the Chern Simons vector potential
and ${\bf A}$ is the vector potential of the external magnetic field;
$\nabla \times \delta{\bf a}=|\Delta_s|^2-|\Delta_{su}|^2$, which
shows that the
variation of pairing wave function results in
$\delta {\bf a}$. ($\Delta_{su}$ is the uniform solution of Eq.6.) 
When $K_s=0$, Eq.6 yields a vortex solution
with the size $\sqrt{\theta_c -\theta_{12}}\xi_0$.
The energy cost of a vortex excitation is 
$g\Delta_0 (\theta_c-\theta_{12})$ for the paired composite Fermion state
coupled with a Chern-Simons field.
Coulomb energies associated with these topological excitations
are negligible because they extend over a distance much larger
than the magnetic length.

\begin{figure}
\begin{center}
\epsfbox{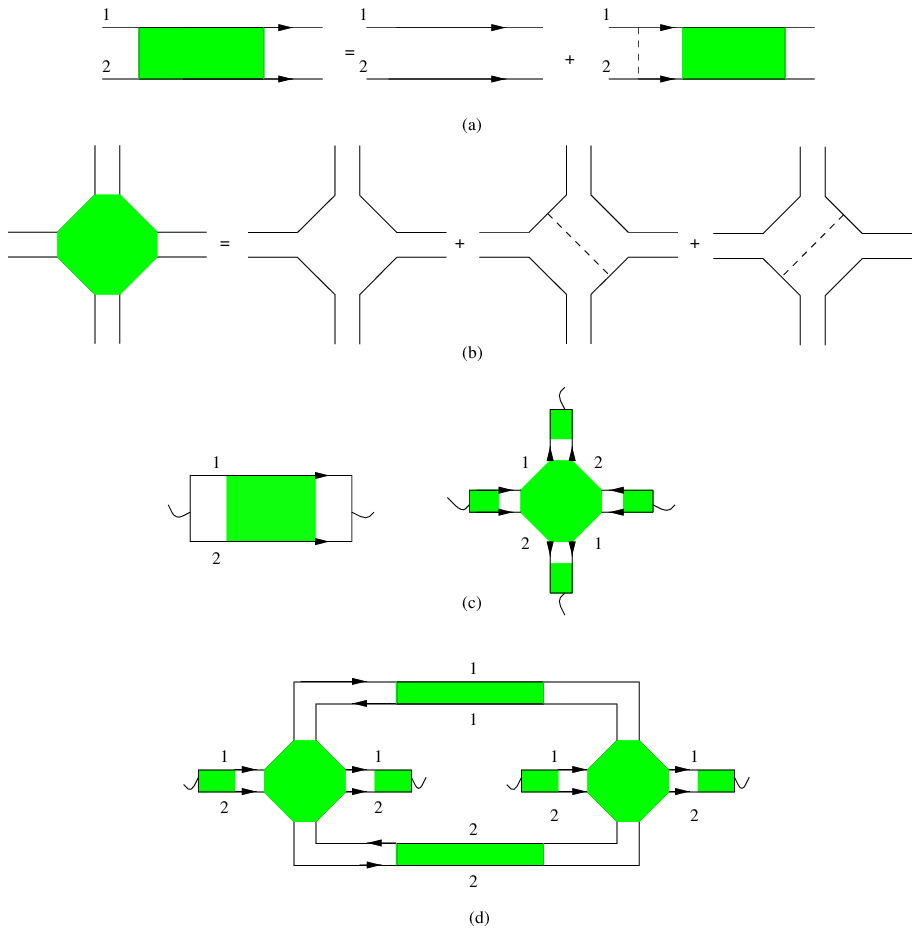}
\leavevmode
\end{center}
\vspace{-5.8cm}
Fig.1. Solid lines are electron Green functions
and dashed lines are impurity scatterings. $1,2$ are layer indices.
Diagrams in b) stand for a Hikami Box.
Diagrams in c) represent the Landau Ginsburg expansion while
the diagram in d) is the typical contribution of $K_s({\bf r}, {\bf r'})$.
\end{figure}

Eq.6 is a stochastic Landau Ginsburge equation with $K_s$ counting
spatial random oscillations of $G_{00}^{1R}G_{00}^{2A}$ at a length
scale much larger than $l/\theta_{12}$.
$1/g^2$ factor in the {\em variance} of 
$K_s({\bf r}, {\bf r'})$ in Eq.6
reflects the Wigner-Dyson
level statistics;
namely, the fluctuation of number of levels within an energy band of Thouless
energy of size of coherence length is unity, regardless 
of the corresponding average number of levels $g$
\cite{Altshuler}.

The statistical property of solutions of this equation
was discussed previously in the context of
a dirty superconducting film in Clogston limit\cite{zhou,Kim}. 
At $\theta_{12} -\theta_c \geq \theta_c/g$,
pairing takes place at a length
scale $L_d=\xi_0\sqrt{|\theta_{12}-\theta_c|/\theta_c}$. The number 
density of these mesoscopic 
states is $1/L^2_d\exp[-g^2|\theta_{12}-\theta_c|/\theta_c]$. 
Most importantly, these mesoscopic states 
are coupled with {\em exchange interactions $\{J\}$
of random signs.}
It belongs to the same universality class as a 
XY model with exchange interactions of random signs.
We restrict ourselves to this limit.

To characterize the ground state and metal stable states, 
I introduce 
the Chern number of the paired state

\begin{eqnarray} 
&&{\cal N}_T= \frac{g\tau^2}{\theta_{12}^2}
\int_{\cal S} d{\bf r}_\alpha \wedge d{\bf r}_\beta [\frac{\partial
\Delta_{s}^{*}({\bf r})}{\partial {\bf r}_\alpha} \frac{\partial
\Delta_{s}({\bf r})}{\partial {\bf r}_\beta} - c.c.]
\nonumber 
\\ &&
=\frac{g\tau^2}{\theta_{12}^2}
\int_{\cal C} d{\bf r} \cdot
Im[\Delta_{s}^{*}({\bf r}) \nabla
\Delta_{s}({\bf r})].
\end{eqnarray} 
The constant in front of the integrals is to ensure
${\cal N}_T$ defined above to be
equivalent
to the standard definition of Chern number for the many body wave
functions $\Psi$, ${\cal N}_T=\sum_{i}
\int dx_i \wedge dy_i [
<\frac{\partial
\Psi}{\partial x_i}| \frac{\partial \Psi}{\partial y_i}> - <\frac{\partial
\Psi}{\partial y_i}|\frac{\partial \Psi}{\partial x_i}>]$.  
$(x_i, y_i)$
are the coordinates of the $ith$ electron.
The last equality in Eq. 7, which follows
Stoke's theory, 
shows that ${\cal N}_T$ is determined by the phase of the condensate
wave function along
${\cal C}$, the boundary of the sample ${\cal S}$.



\begin{figure}
\begin{center}
\epsfbox{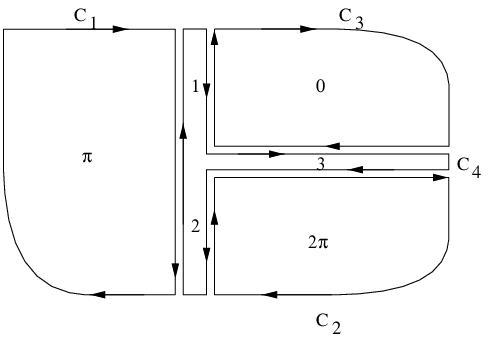}
\leavevmode
\end{center}
\vspace{-0.5cm}
Fig.2. A typical topological excitation. $0, \pi, 2\pi$ stand
for the lift of the condensate phase in those regions.
\end{figure}

First let me point out that ${\cal N}_T$ of the ground state 
is a nonzero sample specific quantity.
Consider a condensate wave function,
which is topologically different from the ground state 
with ${\cal N}'_T \neq {\cal N}_T$.  
Physically, this new condensate wave function
can be obtained by adding a phase $\pi$
to the condensate wave function in the region 
enclosed by contour ${\cal C}_1$, 
$2\pi$ 
phase factor to the region 
enclosed by contour ${\cal C}_2$ while keeping the phase in the region 
enclosed by ${\cal C}_3$ unchanged.
The remaining region is enclosed by a contour ${\cal C}_4$,
which has three extremely narrow stripes, $s_{1,2,3}$.
${\cal N}'_T$ is the summations of four contour integrals
defined in Eq.7 along ${\cal C}_{1,2,3,4}$.

Obviously,
the energy cost of the
new configuration is from 
singular narrow strips $s_{1,2}$, along which the phases
jump by $\pm \pi$. 
Energy of such a configuration is $|J| L^{1/2}$,
with $L^{1/2}$ originating from the summation of {\em random exchange
interaction $\{ J\}$ along the stripes}. 
$|J|$ is a random positive quantity.
Let the probability distribution function be $P(|J|)$, which is continuous
at $J=0$. 
For a given $L$, $L P(J_{min}) J_{min}\approx 1$.  
Therefore $J_{min} \approx L^{-\gamma}$ with $\gamma=1$.
The typical energy of topological excitations 
is proportional to $L^{1/2-\gamma}$. $\gamma \geq 1/2$ means
excitations are of finite energies, as suggested in
\cite{Glass}.

Now let me introduce $\{J'=J+\delta J\}$. 
The energy cost of a configuration with ${\cal N}'_T$ 
considered above
acquires an additional contribution
of a random sign from the contour ${\cal C}_4$, 
of order of $ \delta J L^{1/2}$.
For a given change of $\delta J$, we can always choose a $L$ such that
$J_{min}(L)\propto L^{-\gamma} \ll \delta J$. In this case,
the new ground state belongs to a homotopy class different 
from that of ${\cal N}_T$.
Excitations of different ${\cal N}_T$ with minima energies
of a given homotopy class
correspond to topologically distinct metal stable states.  
The existence of these low energy metal stable states
suggests possible hysteresis. 

Let $\theta=1-<V_1({\bf r})V'_1({\bf r})+V_2({\bf 
r})V'_2({\bf r'})>/<V_1^2({\bf r})+V^2_2({\bf r})>$
stand for the correlation between two different double layer impurity 
potentials $(V_1, V_2)$ and $(V'_1, V'_2)$. 
($<(V'_{1,2})^2>=<(V_{1,2})^2>$.)
Thus, 
${\cal N}_T$ is a nonanalytical function of $\theta$,

\begin{equation}
\lim_{\theta \rightarrow 0}<\frac{[{\cal N}_T(V_1, V_2) 
-{\cal N}_T(V'_1, V'_2)]^2}{{\theta}^2}> =\infty.
\end{equation}
By contrast, the sensitivity of the Chern number defined in 
Eq.8, is zero for a normal mesoscopic sample, a BCS superconductor or
a quantum Hall state.

Infinity sensitivity 
implies that 
there are gapless  
topological excitations. 
To account this, we introduce
\begin{equation}
\frac{\partial {\cal N}_T}{\partial t}=-\frac{{\cal N}_T}{\tau_\kappa},
\end{equation}
with $\tau^{-1}_\kappa$ as the thermal activation rate of all the 
possible topologically distinct excitations. 
The thermal activation rate for a given barrier $E$ is $c T\exp(-E/kT)$,
$c$ is a number of order unity;
the distribution of the finite energy barrier $E$
is assumed to be $P_B(E)$. Naturally, in the presence of
states nearly degenerating with the ground state, $P_B(0) >0$. 
As a result,

\begin{equation}
\frac{1}{\tau_\kappa}=cT\int dV \exp(-\frac{E}{KT})P_B(E)
\approx cT^2 P_B(0),
\end{equation}
which is a power law function of $T$ instead of an exponential
function at low temperature limit. 
As far as $\tau_\kappa$ is longer than any microscopic time scale,
Eq.9 represents the low energy dynamics.

Consequently,
collective modes become strongly damped
because of thermally excited   
topologically different excitations.
The situation here is similar to the gauge theory in the hydrodynamic
limit proposed for Heisenberg spin glass\cite{DV}.
The density-density fluctuation
and current -current fluctuation  
can be calculated as\cite{zhouu}

\begin{eqnarray}
\pi_{dd}=\frac{me^2}{\hbar^2}\frac{v_s^2 q^2}{v_s^2 q^2 +i \omega
\tau^{-1}_\kappa -\omega^2},
\pi_{cc}=\frac{N_se^2}{m}\frac{i\omega}{i\omega -\tau^{-1}_\kappa}.
\end{eqnarray}
which agree with $\pi_{dd}, \pi_{cc}$ obtained
in \cite{zhou99} when topological 
excitations are absent.
$N_s$ and $v_s$ are the superfluid density
and the zero sound velocity respectively in the
absence of topological excitations.
The conductivity of this state is given as

\begin{equation} 
\sigma_{cond}=\lim_{\omega \rightarrow 0, {\bf q}
\rightarrow 0} \frac{i\omega}{q^2}\pi_{dd}(i\omega, {\bf q})= 
\frac{e^2 m}{\hbar^2}
v_s^2 \tau_\kappa(T). 
\end{equation} 
Substituting the results in Eqs.10, 12 into Eq.1, we obtain Eq.3. 
Strictly speaking, at zero temperature $\tau_\kappa^{-1}$ is nonzero,
determined by the quantum tunneling; it is however exponentially small
when $g \gg 1$.

The change of the barrier $\delta E$
in the presence of a transport current $I$ can be estimated according
to minimal coupling principal. 
$\delta E=\hbar/e \int_{\cal S} d{\bf r} {\bf j} \cdot \nabla \chi \sim I
\hbar/e$, with ${\bf j}$ the current density and
$\nabla \chi$ the phase gradient of the condensate.  
When $\delta E$ is of the order of $kT$,
the typical energy of topological excitations, 
the nonlinear effect becomes important. 
This yields a
characteristic current $I_{non} \propto kT\hbar/e^2$,
at which $\sigma_{cond}$ is reduced by a factor of 1/2
of that given in Eq.12.
It implies that $R_{12}(I\gg I_{non}) \ll R_{12}(I=0)$. 
On the other hand, at this current, the
raise of qausiparticle temperature due to Jour heat, $\delta T$, can be 
estimated 
as $\sigma_{quasi}{\bf E}^2 \tau_{ph}/C_v$, with ${\bf E}$ as the
applied
electric
field, $\tau_{ph}$ the inelastic scattering time, $C_v$ the electronic
specific heat.   It is of order of 
$T/g^2 \times L^2_{ph}/L^2$,  
vanishing in the limit when $L_{ph} \ll L$. Here
$L_{ph}$ is the inelastic mean free path, and $L$ is the sample size.

Modulations of local
chemical potentials due to random
impurity potentials lead to 
pinning of vortices as well. 
For short range random impurity potentials with the correlation
length $r_c \ll \xi_0$, 
the pinning potential of a vortex is $\int d{\bf r} 
\delta\rho V_1({\bf r})$
$\propto g \Delta V_0/\epsilon_F \times r_c/\xi_0$; while
the vortex excitation energy in a paired composite
Fermion state is $g\Delta$. 
When $
V_0/\epsilon_F, r_c/\xi_0 \ll 1$, the pinning effect is rather
insignificant.
For the modulation doped {\em GaAs/AlGaAs} 2DEG, 
the correlation length of random impurity potentials
$r_c(\sim 1000 A)$ can be much longer than the Fermi wave length
$k_F^{-1}(\sim 200A)$ and $\xi_0(\sim 200A)$. In this case, 
pinning can be substantial and the coefficient
c in Eq. 10 should be modified. However, the $T^2$ law in Eq.3 remains
valid.

Skyrmions represent
a spin texture which is a nontrivial mapping from a 
sphere onto itself\cite{Sondhi}.
This type excitation has an energy gap due to
Zeeman splitting and Coulomb interaction, which is barely affected by
disorders in the present limit. 
At low temperatures, its contribution
to $\sigma_{cond}$ is negligible.

For a $p$-wave pairing, $\Delta_{pi}$ is suppressed  
in the presence of disorders even when $\theta_{12}=0$
and becomes zero when $\tau=\tau_c\sim 1/\Delta_0$.  
Nontheless, one can make the following substitutions 
in Eq.6 to obtain an equation
for $\Delta_{pi}$ when $\tau$ is close to $\tau_c$:
$\delta\theta_{12}/\theta_c\rightarrow (1-\tau/\tau_c)\delta_{ij}$, 
$\Delta_s \rightarrow \Delta_{pj}$, 
${\tilde{\nabla}}^2\rightarrow 
{\tilde{\nabla}}^2\delta_{ij}+2\tilde{\nabla}_i\tilde{\nabla}_j$,
$K_s({\bf r}, {\bf r'})\rightarrow K_{pij}({\bf r}, {\bf r'})$,
with $<K_{pij}K_{pi'j'}>=[\delta_{ii'}\delta_{jj'}+\delta_{ij}
\delta_{i'j'}]<K_sK_s>$. 
Again, $K_{pij}$ is due to Friedel oscillations
with random phase shifts, i.e. "mismatches" of wave functions
in different oscillation periods.

Though a complete microscopic theory for the pairing quantum
Hall state is still absent, 
it is conceivable that
{\em Eq.3 is qualitatively correct for
other pairing states such as $\{111\}$ state, $\{331\}$ state, or 
Pfaffian states in the presence of
disorders.}
Particularly, the stochastic nature of Eq.6 is generic to a pairing state
involving wave function mismatches.
In weakly coupled limits
the drag conductance is related to the 
topology of the ground state wave function, similar to the
Hall conductance\cite{Yang}.
Recently, I. Aleiner showed that the Coulomb drag 
between two disordered Fermi liquids
can be of a random sign in a mesoscopic limit\cite{Igor}.

I am grateful to 
I. Aleiner, P. W. Anderson, M. Lilly, Y. B. Kim, S. Sondhi, N. Wingreen, and
especially L. Pryadko for many helpful discussions.
I would also like to thank       
NECI, Princeton, USA for its hospitalities.
This work is supported by ARO under DAAG 55-98-1-0270.

\end{multicols}


\begin{thebibliography}{99}
\vspace{-0.3cm}




\bibitem{eisenstein} M. P. Lilly, et.al.,
Phys. Rev. Lett. {\bf 80}, 1714 (1998).
\bibitem{kim} Y. B. Kim, A. J. Millis, preprint, cond-mat/9611125.
\bibitem{stern}I. Ussishkin, A. Stern, Phys. Rev.{\bf B56}, 4013 (1997).
\bibitem{kamenev}A. Kamenev and Y. Oreg, Phys. Rev. B {\bf 52},
7516 (1995).
\bibitem{stern2} I. Ussishkin and A. Stern, preprint, cond-mat/9808013.
\bibitem{zhou99}F. Zhou, Y. B. Kim, Phys. Rev. {\bf B59}, R7825(1999).
\bibitem{Girvin}D. Yoshioka, et.al., Phys. Rev. {\bf B 39}, 1932(1989).
\bibitem{bonesteel} N. E. Bonesteel, Phys. Rev.{\bf B48}, 11484 (1993);
N. E. Bonesteel, et.al., Phys. Rev. Lett. {\bf 77}, 3009(1996). 
\bibitem{Halperin}B. I. Halperin, Helv. Phys. Acta {\bf 56}, 75(1983).
\bibitem{Haldane}F. D. M. Haldane, E. H. Rezayi, Phys. Rev. Lett.
{\bf 60}, 956(1988).
\bibitem{Wen}M. Greiter, et.al., Phys. Rev. Lett.
{\bf 66}, 3205(1991).
\bibitem{Read}G. Moore and N. Read, Nucl. Phys. {\bf B360}, 362(1991).
\bibitem{Read99}T. L. Ho, Phys. Rev. Lett.
{\bf 75}, 1186(1995); N. Read and E. Rezayi, Phys. Rev. 
{\bf B54}, 16864(1996);
N. Read and D. Green, Cond-Mat/9906453.
\bibitem{abr}A.A. Abrikosov, L.P. Gorkov, I.E. Dzyaloshinski,
{\it Methods of Quantum Field Theory in Satistical Physics},
Dover(1961).
\bibitem{Spivak}B. Spivak, A. Zyuzin, 
Chapter 2 in
{\em Mesoscopic
Phenomena in Solids}, edited by B. Altshuler, P.A. Lee and R. Webb,
Elsevier Science Publisher B. V., 1991.
\bibitem{Altshuler}B. Altshuler, B. I. Shklovskii, Zh. Eksp. Teor. Fiz.
{\bf 91}, 220(1986)[Sov. Phys. JETP. {\bf 64}, 127(1986)].
\bibitem{zhou}F. Zhou, B. Spivak, Phys. Rev. Lett. {\bf 80}, 5647(1998).
\bibitem{Kim}I thank Y. B. Kim for pointing out to me
this similarity.
\bibitem{Glass}B.W. Morris, et.al., J. Phys. {\bf C 19}, 1157(1986).
\bibitem{DV}I.E. Dzyaloshinskii, G. E. Volovik,
J. Phys (Paris){\bf 39}, 693(1978).
\bibitem{zhouu}F. Zhou, unpublished, 1999.
\bibitem{Sondhi}S. L. Sondhi, et.al., Phys. Rev. {\bf B47}, 16419(1993).
\bibitem{Yang}K. Yang and A. H. MacDonald, cond-mat/9904061.
\bibitem{Igor}I thank I. Aleiner for sending me 
unpublished results.
\end{thebibliography}
\end{document}